\documentclass[11pt]{article}
\usepackage[hmargin=2.54cm,vmargin=2.54cm]{geometry} 
\usepackage{geometry}
\usepackage{amsmath,amssymb,setspace,fancyhdr,geometry,url,color}
\usepackage{subcaption,graphicx,caption}
\usepackage[authoryear,round]{natbib}
\RequirePackage{lineno} 
\allowdisplaybreaks[0]
\allowbreak

\title{Ice Model Calibration Using Semi-continuous Spatial Data}
\author{Won Chang, Bledar A. Konomi, Georgios Karagiannis, Yawen Guan, Murali Haran}

\newcommand{\balpha}{\ensuremath{\boldsymbol{\alpha}}}

\newcommand{\bdelta}{\ensuremath{\boldsymbol{\delta}}}

\newcommand{\bepsilon}{\ensuremath{\boldsymbol{\epsilon}}}

\newcommand{\bk}{\ensuremath{\mathbf{k}}}
\newcommand{\bK}{\ensuremath{\mathbf{K}}}

\newcommand{\bgamma}{\ensuremath{\boldsymbol{\gamma}}}
\newcommand{\bGamma}{\ensuremath{\boldsymbol{\Gamma}}}

\newcommand{\bmu}{\ensuremath{\boldsymbol{\mu}}}

\newcommand{\boldeta}{\ensuremath{\boldsymbol{\eta}}}

\newcommand{\bpsi}{\ensuremath{\boldsymbol{\psi}}}

\newcommand{\bR}{\ensuremath{\mathbf{R}}}
\newcommand{\bs}{\ensuremath{\mathbf{s}}}

\newcommand{\btheta}{\ensuremath{\boldsymbol{\theta}}}
\newcommand{\bu}{\ensuremath{\mathbf{u}}}

\newcommand{\bv}{\ensuremath{\mathbf{v}}}
\newcommand{\br}{\ensuremath{\mathbf{r}}}
\newcommand{\bw}{\ensuremath{\mathbf{w}}}
\newcommand{\bW}{\ensuremath{\mathbf{W}}}

\newcommand{\bxi}{\ensuremath{\boldsymbol{\xi}}}
\newcommand{\bY}{\ensuremath{\mathbf{Y}}}

\newcommand{\bZ}{\ensuremath{\mathbf{Z}}}

\newcommand{\Cov}{\ensuremath{\mbox{Cov}}}

\newcommand{\bU}{\ensuremath{\mathbf{U}}}

\begin{document}
\maketitle

\begin{abstract}
Rapid changes in Earth's cryosphere caused by human activity can lead to significant
environmental impacts. Computer models provide a useful tool for understanding the behavior
and projecting the future of Arctic and Antarctic ice sheets. However, these models are typically
subject to large parametric uncertainties due to poorly constrained model input parameters that
govern the behavior of simulated ice sheets. Computer model calibration provides a formal
statistical framework to infer parameters using observational data, and to 
quantify the uncertainty in projections due to the uncertainty in these parameters. Calibration
of ice sheet models is often challenging because the relevant model output and observational data
take the form of semi-continuous spatial data, with a point mass at zero and a right-skewed continuous distribution for positive values.  Current calibration approaches cannot handle such data. Here we introduce a hierarchical latent variable model that handles binary spatial patterns and positive continuous spatial patterns as separate components. To overcome challenges due to high-dimensionality we use likelihood-based generalized principal component analysis to impose low-dimensional structures on the latent variables for spatial dependence. We apply our methodology to calibrate a physical model for the Antarctic ice sheet and demonstrate that we can overcome the aforementioned modeling and computational challenges. As a result of our calibration, we obtain improved future ice-volume change projections.  
\end{abstract}


\section{Introduction}
\label{sec:Introduction}
Human-induced climate change is projected to significantly affect the Earth's cryosphere. 
The West Antarctic ice sheet (WAIS) is particularly susceptible to warming climate because a large portion of its body is marine based, meaning that the bottom of the ice is below the sea-level. Any significant changes in this part of Antarctica can lead to a consequential sea level change \citep{fretwell2013bedmap2}. Computer models are used to project the future of WAIS, but the projections from these computer models are highly uncertain due to uncertainty about the values of key model input parameters \citep{stone2010investigating,gladstone2012calibrated,chang2015binary,pollard2016large}. Computer model calibration provides a statistical framework for using observational data to infer input parameters of complex computer models. 

Following the calibration framework described in the seminal paper by \cite{kennedy2001bayesian}, several researchers have developed methods for inferring model parameters for a variety of different types of computer model output. For instance, \cite{bayarri2007computer} provides a wavelet-based approach for calibration with functional model output. \cite{Sanso_forest2009} calibrates a climate model with multivariate output while \cite{Higdon2008} and \cite{chang2013fast} provide approaches for calibrating models with high-dimensional spatial data output. More recently, \cite{chang2015binary} develops an approach for high-dimensional binary spatial data output and \cite{sung2019generalized} proposes a method for binary time series output. \cite{cao2018model} provides a method for censored functional data. Ice sheet thickness data, including the West Antarctic ice sheet data set we consider here, are frequently in the form of high-dimensional semi-continuous spatial data.  No existing calibration methods are suited to this type of data; this motivates the new methodological development in this manuscript.

Several computer model calibration approaches have been applied to infer the parameters and to systematically quantify parametric uncertainty in Antarctic ice sheet models \citep{gladstone2012calibrated,chang2015binary,chang2015improving,pollard2016large,edwards2019revisiting}. One important caveat to existing approaches to ice sheet model calibration is that the model outputs and observational data need to be transformed or aggregated in some degree to avoid issues involving semi-continuous distributions. To be more specific, the main variable of interest in ice model output and observational data is the spatial pattern of ice thicknesses which have positive values at the locations with ice presence and zero values otherwise. Handling such spatially dependent semi-continuous data with truncation at zero poses non-trivial inferential and computational challenges and existing calibration methods cannot readily handle these issues. \cite{chang2015binary} used ice-no ice binary spatial patterns obtained by dichotomizing the thickness patterns into zeros and ones and hence ignored important information regarding the ice thickness. \cite{pollard2016large} also similarly used highly-summarized data to avoid challenges related to semi-continuous data. Although their results show that such approaches still lead to a meaningful reduction in input parameter uncertainty, one can certainly expect that  transforming or summarizing data can result in some significant loss of information. This motivates our methodological development of calibration method that can directly utilize semi-continuous spatial data.

The existing methods for handling semi-continuous data in the spatial statistics literature are based on the truncated Gaussian process approach \citep{stein1992prediction, deoliveira2005bayesian}. In this framework the semi-continuous data being analyzed are viewed as a realization from an underlying Gaussian process, which can be observed only when the values are positive. This simple `clipped' Gaussian process approach provides a natural way to impose spatial dependence among zero and non-zero values. However, the use of truncated process can create serious computational issues when applied to a high dimensional data set with a  large proportion of zeros. This is because  inference based on such a model requires integrating out highly-dependent, high-dimensional, and bounded latent variables for locations with 0 values. Matrix computations for high-dimensional spatial random variables are expensive. Furthermore, designing  efficient (`fast mixing') Markov chain Monte Carlo methods for Bayesian inference for such models becomes very challenging. This is why a clipped Gaussian process \citep[such as one used by][]{cao2018model} does not provide a feasible solution for our calibration problem.


In this paper we formulate an emulation and calibration framework that uses two separate processes: one process for modeling the presence and absence of ice and the other for modeling the value of ice thickness given that ice is present. This approach removes the need to integrate out the bounded latent variables for the locations with no ice and hence allows us to circumvent the related computational challenges in the clipped Gaussian process approach. Our proposed method uses likelihood-based principal component analysis \citep{tipping1999probabilistic} to reduce the dimension of model output and observational data \citep[cf.][]{Higdon2008,chang2013fast}, and avoids issues with large non-Gaussian spatial data calibration \citep[cf.][]{chang2015binary}. In our simulated example and real data analysis, we show that our method can efficiently utilize information from large semi-continuous spatial data and lead to improved calibration results compared to using only binary spatial patterns. While our focus is on calibrating a computer model for the West Antarctic Ice Sheet, the methodology we develop here is readily applicable, with only minor modifications, to other calibration problems with semi-continuous data. 

The rest of this paper is organized as follows. In Section \ref{sec:ModelRunsAndObs}, we introduce the details of our PSU-3D model runs and Bedmap2 observational data that have motivated our methodological development. In Section \ref{sec:EmulationAndCalibration}, we describe our new  framework for emulation and calibration using semi-continuous data and discuss the computational challenges posed by the large size of the spatial data. In Section \ref{sec:DimReduction}, we propose a reduced-dimension approach that can mitigate the computational challenges, and in Section \ref{sec:Results} we describe the result of our analysis on the model runs and observational data using the proposed approach. In Section \ref{sec:Discussion} we summarize our findings and discuss some possible future directions.

\section{Model Runs and Observational Data}
\label{sec:ModelRunsAndObs}
In this study we use a state-of-the-art model, the PSU-3D ice model \citep{pollard2015potential,pollard2016large}, for studying the evolution of WAIS. This model strikes a good balance between model realism and computational efficiency and hence can allow simulations of long term behavior of WAIS (on the scale of thousands of years) with a relatively high resolution of 20 km. Similar to other complex computer model experiments, simulation runs from the PSU-3D ice model are available only at a limited number of input parameter settings due to the high computational cost. Therefore in this study we take an emulation approach in which we first create a collection of model runs at pre-specified design points in the input parameter space (often called a perturbed physics ensemble) and then build a statistical surrogate based on those model runs. 

We use a previously published ensemble of simulations  \citep{chang2015improving,pollard2016large} generated from PSU-3D ice model with 625 model runs. The parameter settings for ensemble members are determined by a factorial design with 5 design points for each input parameter. There are four input parameters varied in the ensemble: sub-ice-shelf oceanic melt factor (OCFAC, non-dimensional), which determines  oceanic melting at the bottom of floating ice shelves caused by the changes in the surrounding ocean temperature; calving factor (CALV, non-dimensional), the rate of calving of iceberg at the oceanic edge of floating shelves; basal sliding coefficient (CRH, m year$^{-1}$ Pa$^{-2}$); velocity of sliding movement of grounded ice, determined by the interface between the grounded ice and its bed rock; asthenospheric relaxation e-folding time (TAU, 1000 years), the time scale for ice sheet evolution caused by changing ice load on its bedrock. While these parameters play important roles in determining the long-term evolution of the Antarctic ice sheet, their values are highly uncertain and hence need to be properly calibrated for realistic simulation.

\begin{figure}
\centering
\includegraphics[scale=0.7]{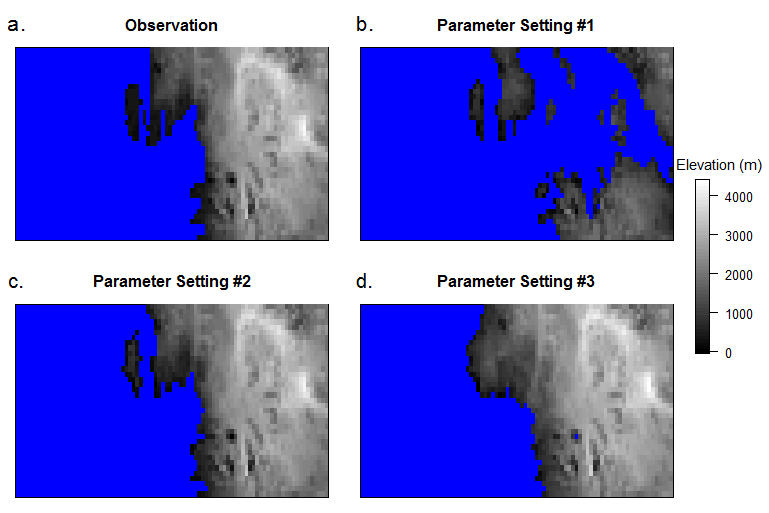}
\caption{Observational data (a) from Bedmap 2 data \citep{fretwell2013bedmap2} and example model runs (b-d) from PSU-3D ice model.}
\label{fig:ModelAndData}
\end{figure}

Each ensemble member is spun up from 40,000 years before present to modern times and then projected into future for 5,000 years. We then extract the spatial patterns of modern grounded ice sheet thickness in Amundsen Sea Embayment (ASE) region, which is expected to be one of the major contributor to sea level change in the future. The spatial pattern in our selected region has $86\times37$ pixels with 20 km $\times$ 20 km resolution (Figures \ref{fig:ModelAndData} b-d). To calibrate the four input parameters introduced above, we compare these model outputs with the observed modern ice sheet thickness pattern in the same area derived from the Bedmap2 dataset \citep{fretwell2013bedmap2} (Figure \ref{fig:ModelAndData} a). This recent data product combines a wide range of sources including seismic sounding, radar surveys, and satellite altimetry. Since the observational grid has a higher spatial resolution (1 km $\times$ 1 km resolution), we upscale the observational data to the model grid using a simple linear interpolation. Note that the model outputs and the observational data for ice thickness are all in the form of high-dimensional semi-continuous spatial data which poses non-trivial statistical challenges for our calibration framework.

\section{Computer Model Emulation and Calibration Using Semi-continuous Spatial Data}
\label{sec:EmulationAndCalibration}
In this section we describe our statistical framework for inferring the input parameters in the PSU-3D ice model. In particular we focus on describing how the standard computer model emulation and calibration framework \citep{kennedy2001bayesian} can be modified to accommodate the ice thickness patterns introduced above, which take the form of semi-continuous data. 


We use the following notation hereafter: Let the $p$-dimensional vector $\bY(\btheta)=[Y(\btheta,\bs_1)$, $\dots$,$Y(\btheta,\bs_{p})]^T$ denote the spatial pattern of ice thickness at the spatial locations of the model grid $\bs_1,\dots,\bs_{p} \in R^2$ which is generated from the computer model given input parameter setting $\btheta \in R^d$.  Here, $d$ is the dimension of the input space which in our application is equal to four. The observed data at the same spatial locations are denoted as a  $p$-dimensional vector    $\bZ=[Z(\bs_1),\dots,Z(\bs_{p})]^T$.  Here, $Y(\btheta,\bs_j)$ and   $\bZ(\bs_j)$ can have either positive values representing the ice thickness or zero values denoting absence of ice  at location $\bs_j$ (see Figures \ref{fig:ModelAndData}). 



We denote the design points for the input parameters in our ensemble as $\btheta_1,\dots,\btheta_{n}$. As a result the collection of model output in our ensemble can be denoted as an $n\times p$ matrix $\bY$, with elements  $\left[\bY\right]_{i,j} = Y(\btheta_{i},\bs_{j})$ for $i=1,...,n$ and $j=1,...,p$, where the rows correspond to different input parameter settings while the columns correspond to different spatial locations. In our ice thickness application the number of spatial locations for the grid is $p=86\times37=3,182$ and the number of model runs in the ensemble is $n=625$.

\subsection{Computer Model Emulation Using Semi-Continuous Spatial Data}
\label{sec:EmulationRuns}

Since only a limited number of computer runs can be carried out, we use an emulator to statistically link the modeled ice thickness to the observational data.  However, the semi-continuous nature of $\bY(\btheta)$ prevents direct application of existing GP calibration approaches such as those in \cite{sacks1989design} and \cite{kennedy2001bayesian}. In order to make emulation of the semi-continuous $Y(\btheta,\bs_j)$ variable possible, we introduce an indicator variable $I_y(\btheta_i,\bs_j)$ whose value is one if grounded ice is  present at the given parameter setting and spatial location $(\btheta_i,\bs_j)$ or zero otherwise for $i=1,\dots,n$ and $j=1,\dots,p$. Given that grounded ice is present, we model the thickness as $Y(\btheta_i,\bs_j)=q\left( h(\btheta_i,\bs_j) \right)$, where  $q:\mathbb{R}\rightarrow \mathbb{R}_{+}$ is a bijective transformation function that allows $h(\btheta_i,\bs_j)$ to take any real  value. We can now formulate the ice thickness $Y(\btheta_i,\bs_j)$ as 
                                                                 
\begin{equation} \label{eqn:YtwoWay}
Y(\btheta_i,\bs_j)=
\begin{cases}
    q\left( h(\btheta_i,\bs_j) \right) &,~ \mbox{if}~ I_y (\btheta_i,\bs_j)=1 \\
    0 &, ~\mbox{if}~ I_y (\btheta_i,\bs_j)=0
  \end{cases},
\end{equation}
for $i=1,\dots,n$ and $j=1,\dots,p$.  Using this representation we can translate the problem of emulating $\bY(\btheta)$ into the problem of finding the predictive distributions of the binary response $I_y(\btheta,\bs_1),\dots,I_y(\btheta,\bs_{p})$ and the transformed thickness values  $\mathbf{h}(\btheta)=\left[h(\btheta,\bs_1),\dots,h(\btheta,\bs_{p})\right]^T$ at any untried input parameter setting $\btheta$. Therefore, we can model $\mathbf{h}(\btheta)$ directly as a multivariate Gaussian process, since its elements are unbounded and continuous. We use a $p$-dimensional vector $\boldeta(\btheta)=\left[\eta(\btheta,\bs_1),\dots,\eta(\btheta,\bs_{p})\right]^T$ to denote the emulated process for $\mathbf{h}(\btheta)$ from a GP. For the binary spatial pattern $I_y(\btheta,\bs_1),\dots,I_y(\btheta,\bs_{p})$, we indirectly emulate them through their corresponding logits  $\bgamma(\btheta)=[\gamma(\btheta,\bs_1),\dots,\gamma(\btheta,\bs_{p})]^T$ defined as 

\begin{equation*} 
P(I_y(\btheta,\bs_j)=x)=\left(\frac{\exp(\gamma(\btheta,\bs_j))}{1+\exp(\gamma(\btheta,\bs_j))}\right)^x\left(\frac{1}{1+\exp(\gamma(\btheta,\bs_j))}\right)^{1-x},
\end{equation*}
for $j=1,\dots,p$ as in \cite{chang2015binary}. Since $\boldsymbol{\gamma}(\btheta)$$=[\gamma(\btheta,\bs_1)$ $,\dots,\gamma(\btheta,\bs_{p})]^T$ can be again treated as continuous variables with unbounded support the use of the standard GP approach is suitable. Since $\gamma(\btheta,\bs_j)$ is an unobserved latent variable even if $\btheta$ is one of the existing design points $\btheta_1,\dots,\btheta_n$, we do not use a separate notation for the logits at those design points. Our emulation problem now becomes a problem of finding predictive processes $\boldeta(\btheta)$ and $\bgamma(\btheta)$ at any untried settings $\btheta$ (which are possibly dependent to each other).

\subsection{Computer Model Calibration Using Semi-Continuous Spatial Data} \label{sec:CalibrationFramework}


Formulating calibration framework also requires to address the issue with semi-continuous data because the standard calibration approach \citep{kennedy2001bayesian} is not applicable. Here we use a similar representation of the observed ice thickness $Z(\bs_j)$ as in \eqref{eqn:YtwoWay}. We define the variable $I_z(\bs_j)$ to be an indicator with a value of one if observed grounded ice presents at $\bs_j$ and zero otherwise. To transform the observational data, we use the same transformation function $q$ as in  \eqref{eqn:YtwoWay}. 
At any spatial location  $\bs_j$, we assume observation of ice thickness $Z(\bs_j)$ can be represented as follows:
\begin{equation} \label{equation:TwoWayForZ}
Z(\bs_j)=
\begin{cases}
    q\left( t(\bs_j)\right) &,~ \mbox{if}~  I_z(\bs_j) =1 \\
    0 &, ~\mbox{if}~   I_z(\bs_j) =0
  \end{cases}.
\end{equation}
In a similar fashion to our emulation framework, we set up our model for the transformed thickness $t(\bs_j)$ and the logit of $I_z(\bs_j)$ denoted as $\lambda(\bs_j)$. Following \cite{chang2015binary}  we set up the following model to link it to the logit for the model output at the best setting ($\gamma(\btheta^*,\bs_j)$) while accounting for data-model discrepancy:
\begin{equation}  \label{equation:LogitForZ}
\lambda(\bs_j) =\gamma(\btheta^*,\bs_j)+ \alpha(\bs_j),
\end{equation}
where $\btheta^*$ is the input parameter setting that gives the `best' match between model output and observational data and a normal random variable $\alpha(\bs_i)$ is a spatially correlated discrepancy term. 

The model for $t(\bs_j)$ needs to be defined only for the locations with $I_z(\bs_j) =1$. Let $m=\sum_{j=1}^{p} I_z(\bs_j)$ be the number of spatial locations with a positive observed thickness. Without loss of generality, we assume that the observed thicknesses at the first $m$ locations $\bZ^{+}=[Z(\bs_1),\dots,Z(\bs_m)]$ are positive while the rest $Z(\bs_{m+1}),\dots,Z(\bs_p)$ are $0$. For $\bs_1,\dots,\bs_m$, we use the following model for the transformed thickness: 
\begin{equation} \label{equation:ZetaModel}
t(\bs_j)=\eta(\btheta^{*},\bs_j)+\delta(\bs_j)+\epsilon(\bs_j),
\end{equation}
where the random variables $\bdelta=[\delta(\bs_1),\dots,\delta(\bs_{m})]^T \sim N(\mathbf{0},\Sigma_\delta)$ and $\bepsilon=[\epsilon(\bs_1),\dots,\epsilon(\bs_{m})]^T$ $\sim N(\mathbf{0},\sigma_\epsilon^2 \mathbf{I}_m) $ respectively represent the spatially correlated data-model discrepancy and the i.i.d. observational error. The discrepancy covariance $\Sigma_\delta$ reflects the spatial dependence among $\delta(\bs_1),\dots,\delta(\bs_{m})$.

 The model in \eqref{equation:LogitForZ} assigns the following Bernoulli distribution for $I_z(\bs_j)$ (conditionally on the value of $\btheta^*$ and discrepancy $\alpha(\bs_j)$):{\small
\begin{equation*} 
P(I_z(\bs_j)=x|\gamma(\btheta^*,\bs_j),\alpha(\bs_j))=\left(\frac{\exp(\gamma(\btheta^*,\bs_j)+\alpha(\bs_j))}{1+\exp(\gamma(\btheta^*,\bs_j)+\alpha(\bs_j))}\right)^x \left(\frac{1}{1+\exp(\gamma(\btheta^*,\bs_j)+\alpha(\bs_j))}\right)^{1-x}.
\end{equation*}}
Given this distribution for $I_z(\bs_j)$, we can view the specification in \eqref{equation:TwoWayForZ} 
 as a mixture model with the following density:
\begin{equation}
\begin{aligned} \label{equation:ZbasicMixture}
  f( Z(\bs_j)| \eta(\btheta^{*},\bs_j),&\delta(\bs_j),\sigma^2_\epsilon, \gamma(\btheta^*,\bs_j),\alpha(\bs_j)) \\
    = &\left| \frac{\partial Z(\bs_j) }{\partial t(\bs_j)} \right| f\left(t(\bs_j)|\eta(\btheta^{*},\bs_j),\delta(\bs_j),\sigma^2_\epsilon\right)P\left(I_z(\bs_j)=1|\gamma(\btheta^*,\bs_j),\alpha(\bs_j)\right)\\ &\qquad+\mathcal{D}_0\left(Z(\bs_j)\right)P\left(I_z(\bs_j)=0|\gamma(\btheta^*,\bs_j),\alpha(\bs_j)\right) 
\end{aligned} 
\end{equation}
for all locations $\bs_1,\dots,\bs_p$, where the density function  $f\left(t(\bs_j)|\eta(\btheta^{*},\bs_j),\delta(\bs_j),\sigma^2_\epsilon\right)$ is given by \eqref{equation:ZetaModel} and $\mathcal{D}_0$ is the direct delta function. Since the density in \eqref{equation:ZbasicMixture} can be re-written as 
\begin{equation*}
\begin{aligned}
&f\left(t(\bs_j)|\eta(\btheta^{*},\bs_j)\right.,\left.\delta(\bs_j),\sigma^2_\epsilon  \right)\\
&=\begin{cases}
     f\left(t(\bs_j)|\eta(\btheta^{*},\bs_j),\delta(\bs_j),\sigma^2_\epsilon\right) P(I_z(\bs_j)=1|\gamma(\btheta^*,\bs_j),\alpha(\bs_j)) &,~ \mbox{if}~   I_z(\bs_j) =1, \\
     P(I_z(\bs_j)=0|\gamma(\btheta^*,\bs_j),\alpha(\bs_j))  &,~\mbox{if}~   I_z(\bs_j) =0,
  \end{cases}
\end{aligned}
\end{equation*}
and $Z(\bs_1),\dots,Z(\bs_p)$ are conditionally independent given the relevant parameters, the likelihood for $\bZ$ can be factorized as follows:
\begin{equation}
\begin{aligned} \label{equation:FactoredLikelihoodForZ}
\mathcal{L}&( \bZ|\boldeta^+(\btheta^{*}),\bdelta,\sigma^2_\epsilon, \bgamma(\btheta^*),\balpha)
\\
\propto& \prod_{j=1}^{m}  f\left(t(\bs_j)|\eta(\btheta^{*},\bs_j),\delta(\bs_j),\sigma^2_\epsilon\right) P(I_z(\bs_j)=1|\gamma(\btheta^*,\bs_j),\alpha(\bs_j)) \\
 &\times \prod_{j=m+1}^{p} P(I_z(\bs_j)=0|\gamma(\btheta^*,\bs_j),\alpha(\bs_j)),\\
=&\mathcal{L}_1\left(\bZ^{+}|\boldeta^+(\btheta^{*}),\bdelta,\sigma^2_\epsilon\right) \mathcal{L}_2(I_z(\bs_1),\dots, I_z(\bs_p)| \bgamma(\btheta^*),\balpha),
\end{aligned} 
\end{equation} 
where 
\begin{align*}
\mathcal{L}_1\left(\bZ^{+}|\boldeta^+(\btheta^{*}),\bdelta,\sigma^2_\epsilon\right)=&\prod_{j=1}^{m}  f\left(t(\bs_j)|\eta(\btheta^{*},\bs_j),\delta(\bs_j),\sigma^2_\epsilon\right),\\
\mathcal{L}_2(I_z(\bs_1),\dots, I_z(\bs_p)|\bgamma(\btheta^*),\balpha)=&\prod_{j=1}^{m} P(I_z(\bs_j)=1|\gamma(\btheta^*,\bs_j),\alpha(\bs_j)) \\
&\times \prod_{j=m+1}^{p}  P(I_z(\bs_j)=0|\gamma(\btheta^*,\bs_j),\alpha(\bs_j)).
\end{align*} 
Here $\boldeta^+(\btheta^{*})$ is the vector of emulated process for all positive $Z(\bs_j)$'s (i.e. $\boldeta^+(\btheta^{*})=[\eta(\btheta^{*},\bs_1),\cdots,\eta(\btheta^{*},\bs_m)]^T$), and $\balpha=[\alpha(\bs_1),\dots,\alpha(\bs_p)]^T$. The Jacobian factors $\left| \frac{\partial Z(\bs_j) }{\partial t(\bs_j)} \right|$ are omitted as they do not depend on any model parameters. 

Note that this formulation does not necessarily require independence between $\bZ^{+}$ and $I_z(\bs_1),\dots,I_z(\bs_p)$, because dependence can easily be specified through dependence between $\eta(\btheta^{*},\bs_j)$ and $\gamma(\btheta^{*},\bs_j)$ or $\delta(\bs_j)$ and $\alpha(\bs_j)$. This is how we impose dependence between $\bZ^{+}$ and $I_z(\bs_1),\dots,I_z(\bs_p)$ in our formulation (see Section \ref{section:CalibrationBasis} below).

The factorization above shows that the likelihood for $\bZ$ can be factored into two parts, one for the positive observations $\bZ^{+}$ and the other for the indicator variables at all locations $I_z(\bs_1),\dots, I_z(\bs_p)$. This has an important implication for inference on $\btheta^*$: utilizing the ice thickness pattern for calibration is essentially using the additional information from the positive ice thickness values $\bZ^+$ on top of the binary spatial pattern of ice presence ($I_z(\bs_1),\dots,$ $I_z(\bs_p)$) in calibration. We will show  how this added information improves our inference on the input parameter $\btheta^*$ in both the simulated and the real data examples in Section \ref{sec:Results} below.

\subsection{Computational and Inferential Challenges}
\label{sec:Challenges}

The basic framework described in the previous section faces some computational and inferential challenges when the model output and the observational data are in the form of high-dimensional spatial data (i.e, $p$ is large) as in our PSU-3D Ice model calibration problem: First, inference based on the formulations described in Sections \ref{sec:EmulationRuns} and \ref{sec:CalibrationFramework} requires to handle a large number of latent variables for the logits. To be more specific the number of latent variables in the emulation step is $n\times p$ and this translates to $625\times3,182\approx$2 million variables to infer for our problem. In the calibration step, while the number of latent variables is much smaller  than that in the emulation step ($2p=6,364$), the number of available data points ($p$) is much smaller than the number of latent variables ($2p$) and hence the problem is in fact ill-posed. Second, the size of data for height patterns from the model output is still large even when we consider only those at $\btheta_i$ and $\bs_j$ with $I_y(\btheta_i,\bs_j)=1$. In our calibration problem, the number of $(\btheta_i,\bs_j)$ combinations with $I_y(\btheta_i,\bs_j)=1$ is about 8.5 million and this makes the standard Gaussian process emulation approach computationally infeasible because of the well-known computational issue with a large covariance matrix \citep[see, e.g.,][]{heaton2018case}. Due to the highly irregular spatial distribution of the locations with $I_y(\btheta_i,\bs_j)=1$ a simple solution such as using a separable covariance structure \citep[see, e.g.,][]{gu2016parallel} is not applicable here.

\section{Dimension Reduction-Based Approach}
\label{sec:DimReduction}
We mitigate the aforementioned challenges due to high-dimensional spatial data using the likelihood-based principal component analysis (PCA) methods \citep{tipping1999probabilistic}. Unlike the singular value decomposition-based PCA, the likelihood-based PCA can easily handle non-Gaussian data or partially observed data and hence is highly suitable for our problem. 

\cite{salter2019uncertainty} recently has cautioned about possible issues regarding use of principal components in calibration--if the overall range for model output does not cover the range for observational data, calibration based on principal components can yield nonsensical results. \cite{salter2019uncertainty} has also proposed an optimal basis approach that can provide a solution in such situation. \cite{chang2013fast} and \cite{chang2015improving} also discuss possible issues in a similar vein from the viewpoint of constructing discrepancy terms. Since the model runs and observational data discussed in Section \ref{sec:ModelRunsAndObs} does not have such issues, we choose not to implement the optimal basis approach by \cite{salter2019uncertainty}.

\subsection{Emulation Based on Likelihood-based Principal Component Analysis.}~  \label{subsection:likelihoodPCA}
Let $\boldsymbol{\Gamma}=\left[\boldsymbol{\gamma}\left(\btheta_1\right),\dots,\boldsymbol{\gamma}\left(\btheta_{n}\right)\right]^T$ be a matrix of logits for the binary patterns $\left\{I_y(\btheta_i,\bs_j) \right\}$ ($i=1,\dots,n$ and $j=1,\dots,p$) for the existing model runs. The rows of $\bGamma$ correspond to the design points in input parameter settings $\btheta_1,\dots,\btheta_{n}$ while the columns are for different spatial locations $\bs_1,\dots,\bs_{p}$. We  apply logistic principal component analysis \citep{lee2010sparse} to decompose the logit matrix $\bGamma$ in the following way:
\begin{equation} \label{equation:matrix_dim_reduction}
\boldsymbol{\Gamma}=\mathbf{1}_n  \boldsymbol{\mu}^T + \bW \bK_w^T,
\end{equation}
where $\bmu$ is the $p\times 1$ mean vector for the spatial locations $\bs_1,\dots,\bs_{p}$ (i.e. the column means of $\bGamma$), $\bW$ is the $n\times J_w$ logistic principal component (LPC) score matrix, and $\bK_w$ is the $p \times J_w$ LPC matrix with a pre-specified number of principal components $J_w\ge 1$. The rows of $\bW=\left[ \mathbf{w}(\btheta_1), \dots,\mathbf{w}(\btheta_{n}) \right]^T$ correspond to the logits for different input parameter settings where $\mathbf{w}(\btheta)=[w_1(\btheta),\dots w_{J_w}(\btheta) ]^T$ denotes a vector of the LPC scores at $\btheta$. The parameters in \eqref{equation:matrix_dim_reduction} ($\boldsymbol{\mu}$, $\bW$, and $\bK_w$) can be estimated by maximizing the corresponding likelihood function for these parameters given the binary patterns $\left\{I_y(\btheta_i,\bs_j) \right\}$ for existing model runs using the minorization and maximization (MM) algorithm.  We predict the logits $\boldsymbol{\gamma}(\btheta)$ at any untried setting $\btheta$ by predicting the corresponding LPC scores $\mathbf{w}(\btheta)$. 

Each score $w_k(\btheta)$ (for $k=1,\dots J_w$) can be predicted separately using a GP emulator with a zero mean and the following exponential covariance function.
\begin{equation*}
\Cov(w_k(\btheta),w_k(\btheta'))=\zeta_{w,k} I(\btheta = \btheta') + \kappa_{w,k} \exp\left(-\sum_{b=1}^d \frac{|\theta_b-\theta'_b|}{\phi_{w,kb}}\right)
\end{equation*}
for two possibly different input parameter settings $\btheta$ and $\btheta'$ where $\zeta_{w,k}>0$ is the nugget, $\kappa_{w,k}>0$ is the partial sill, and $\phi_{w,k1},\dots,\phi_{w,kd}>0$ are the range parameters. We find the maximum likelihood estimates of the covariance parameters $\hat{\zeta}_{w,k}$, $\hat{\kappa}_{w,k}$, and $\hat{\phi}_{w,k1},\dots,\hat{\phi}_{w,kd}$ to construct an emulator for individual principal components. We denote the resulting emulated process of LPC scores at $\btheta$ as $\boldsymbol{\psi}(\btheta)=[\psi_1(\btheta),\dots,\psi_{J_w}(\btheta)]^T$. 

We also apply a likelihood-based PCA method for data with missing values to build an emulator for the ice-thickness patterns. For $\btheta_i$ and $\bs_j$ with $I_y(\btheta_i,\bs_j)=1$ we assume the following model for dimension reduction:
\begin{equation*}
h(\btheta_i,\bs_j)= \sum_{l=1}^{J_{u}} k_{u,jl} u_l(\btheta_i) + e_{ij} 
\end{equation*}
with $e_{ij} \sim \mbox{i.i.d} ~N(0,\sigma_e^2)$ ($\sigma_e^2>0$), the principal component (PC) loading $ k_{u,jl}$ ($j=1,\dots,p$ and $l=1,\dots,J_u$) and the PC score $u_l(\btheta_i)$ ($i=1,\dots,n$ and $l=1,\dots,J_u$). Again $J_u\ge 1$ is the pre-determined number of principal components being used for our dimension reduction. This is essentially PCA with missing values and therefore the PC loadings and scores can be estimated via EM algorithm \citep{stacklies2007pcamethods}. We denote the resulting $p \times J_u$ loading matrix by $\bK_u$, with $(i,j)$th element given by $k_{u,ij}$. In a similar manner to the problem of emulating logits we predict the latent variables for the thickness $h(\btheta,\bs_j)$ at any untried setting $\btheta$ and location $\bs_j$ with a positive thickness value by predicting the corresponding principal component scores $\mathbf{u}(\btheta)=[u_1(\btheta),\dots,u_{J_u}(\btheta)]^T$.

Again we build an emulator for each principal component separately using a GP emulator with the following exponential covariance function:
\begin{equation} \label{equation:covariance_for_u}
\Cov(u_l(\btheta),u_l(\btheta'))=\zeta_{u,l} I(\btheta = \btheta') + \kappa_{u,l} \exp\left(-\sum_{b=1}^d \frac{|\theta_b-\theta'_b|}{\phi_{u,lb}}\right)    
\end{equation}
for any two input parameter settings $\btheta$ and $\btheta'$ where $\zeta_{u,l}>0$ is the nugget, $\kappa_{u,l}>0$ is the partial sill, and $\phi_{u,l1},\dots,\phi_{u,ld}>0$ are the range parameters. To incorporate information from the binary pattern we use the following mean function for the $l$th principal component: 
\begin{equation} \label{equation:EmulMean}
E\left(u_l(\btheta_i)|w_1(\btheta_i),\dots,w_{J_w}(\btheta_i)\right)=\sum_{k=1}^{J_{w}}  g_{lk} (w_k(\btheta_i)),
\end{equation}
where the function $g_{lk}$ is given by a natural spline regression model whose degrees of freedom is determined through cross-validation \citep{hastie1992generalized}. We let $\boldsymbol{\beta}_{lk}$ be the vector of coefficients for $g_{lk}(\cdot)$, whose dimensionality is the same as the degrees of freedom of $g_{lk}$. To construct the GP emulator we find the estimates of the covariance parameters (denoted as $\hat{\zeta}_{u,l}$, $\hat{\kappa}_{u,l}$ and $\hat{\phi}_{u,l1},\dots,\hat{\phi}_{u,ld}$) and the parameters for the spline functions (denoted as $\hat{\boldsymbol{\beta}}_{l1},\dots,\hat{\boldsymbol{\beta}}_{lJ_w}$) for each $l$th principal component separately via restricted maximum likelihood estimation (REML) \citep{stein1999interpolation}. When we predict $u_l(\btheta)$ for any untried setting $\btheta \notin \left\{\btheta_1,\dots,\btheta_{n} \right\}$, we replace $w_k(\btheta)$  with  $E(\psi_k(\btheta)|w_k(\btheta_1),\dots,w_k(\btheta_{n}))$ given by the Gaussian process emulator described above since $w_k(\btheta)$ is not available if $\btheta \notin \left\{\btheta_1,\dots,\btheta_{n} \right\}$. We let $\boldsymbol{\xi}(\btheta)=[\xi_1(\btheta),\dots,\xi_{J_u}(\btheta)]^T$ denote the resulting emulated process for $\mathbf{u}(\btheta)$.

For any untried input parameter setting $\btheta$, we can predict the ice thickness pattern from our computer model in the following two steps: (i) We first predict the logits of ice-no ice patterns  $\bgamma(\btheta)$ as $\bK_w \bpsi(\btheta)$, and (ii) for each location $\bs_j$ with $\gamma(\btheta,\bs_j)>0$ the predicted thickness is given as $q\left(\sum_{l=1}^{J_{u}} k_{u,jl} u_l(\btheta)\right)$. Note, however, that the thresholding of the logits at 0 is needed only for evaluating emulation performance (such as generating predicted patterns for visual evaluation) and is not used in our actual calibration procedure.

In the calibration step discussed below, we fix the emulator parameters at their MLEs except for the partial sill parameters for $\bxi$, $\boldsymbol{\kappa}_u=\left[ \kappa_{u,1},\dots,\kappa_{u,J_u} \right]$.  The partial sill parameters for $\bxi$ will be re-estimated along other parameters in the calibration model to account for any possible discrepancies in scale \citep[see e.g.,][for smiliar approaches ]{bhat2013inferring,chang2013fast,chang2013composite,chang2015improving}. However, the partial sills for $\bpsi$ will be fixed at their MLEs without being re-estimated in the calibration stage because the binary patterns usually do not have enough information for the scale parameters of the latent variables and hence re-estimation for the partial sill parameters often cause identifiability issues as discussed in \cite{chang2015binary}.

\subsection{Calibration Using Basis Representation}
\label{section:CalibrationBasis}
Using the emulators for principal components ($\bpsi$ and $\bxi$) described in the previous section we modify the basic calibration framework introduced in Section \ref{sec:CalibrationFramework} to set up a computationally efficient calibration method. We now rewrite the model for $t(\bs_j)$ in \eqref{equation:ZetaModel} as
\begin{equation} \label{equation:BasisRepForZ}
t(\bs_j)=\sum_{l=1}^{J_{u}} k_{u,jl} \xi_l(\btheta^*) +  \sum_{k=1}^{J_{r}} k_{r,jk} r_k+ \epsilon_j
\end{equation}
for $j=1,\dots,m$, where $k_{r,jk}$ is the $(j,k)$th element of an $m \times J_{r}$ discrepancy basis matrix $\mathbf{K}_r$, $r_1,\dots,r_{J_{r}}\sim i.i.d. ~N(0,\sigma_{r}^2)$ are the random coefficients with $\sigma_{r}^2>0$ for $\mathbf{K}_r$, and $\epsilon_j\sim~N(0,\sigma_\epsilon^2)$ is the i.i.d. observational error with $\sigma_\epsilon^2>0$. The terms $\sum_{l=1}^{J_{u}} k_{u,jl} \xi_l(\btheta^*)$ and $\sum_{k=1}^{J_{r}} k_{r,jk} r_k$ are  respectively the basis representations of $\eta(\btheta^*,\bs_j)$ and $\delta(\bs_j)$ in \eqref{equation:ZetaModel} given by our formulation. We also rewrite the model for the logits $\boldsymbol{\lambda}$ for $\bZ$ in \eqref{equation:LogitForZ} using a similar basis representation as follows:
\begin{equation} \label{equation:BasisRepForLambda}
\boldsymbol{\lambda} =\boldsymbol{\mu}+ \bK_{w} \boldsymbol{\psi} \left(\btheta^*\right)+ \bK_v \bv,
\end{equation}
with the logistic principal component basis matrix $\bK_{w}$, and a $p \times J_v$ discrepancy basis matrix $\bK_v$ and its corresponding coefficients $\bv = [v_1,\dots,v_{J_v}]^T \sim N(0,\sigma_{v}^2 I_{J_v})$ with $\sigma_{v}^2 >0$. We model the dependence between the coefficients of the discrepancy terms $\bv=[v_1,\dots,v_{J_v}]^T$ and $\mathbf{r}=[r_1,\dots,r_{J_r}]^T$ through a $J_v \times J_r$ cross correlation matrix $\mathbf{R}$, whose $(i,j)$th element $\rho_{ij}$ is the correlation between $v_i$ and $r_j$.

The discrepancy basis matrices $\bK_r$ and $\bK_v$ need to be carefully specified to avoid possible identifiability issues between the effects of input parameters and the discrepancy. For the discrepancy basis for the thickness $\bK_r$ we use the kernel convolution \citep{higdon1998process,Higdon2008} with 40 knots that are  evenly distributed in the spatial domain, with the exponential kernel function with the range parameter of 400 km. To reduce the  identifiablity issues we use the 10 leading eigenvectors of the kernel matrix as $\bK_r$ instead of the original kernel matrix. Using eigenvectors instead of the original the basis matrix has a similar regularizing effect as a ridge regression \citep{trevor2009elements}. Similarly found in \cite{chang2013fast}, our pilot simulation study shows that the value of the range parameter for the kernel function has very minimal effect on the inference result (results not shown). For the discrepancy basis matrix for the binary pattern $\bK_v$ we use the data-driven basis described in \cite{chang2015binary}. To be more specific for each spatial location $\bs_j$  we compute the following measure of signed mismatch between the model output and observational data:
\begin{equation*}
\mbox{dif}_j=\frac{1}{n}\sum_{i=1}^n  (I_y(\btheta_i,\bs_j)-I_z(\bs_j)).
\end{equation*}
The $j$th element of the basis vector $\bK_v=\bk_{v}$ (i.e. $J_v$ is set to be 1) is then defined as 
\begin{equation*}
\bk_{v,j}=
\begin{cases}
    \log\left( \frac{1+\mbox{dif}_j}{1-\mbox{dif}_j}\right) &,~ \mbox{if}~ \left| \mbox{dif}_j \right|>0.5, \\
    0 &, ~\mbox{if}~   \left| \mbox{dif}_j \right| \le 0.5.
  \end{cases}
\end{equation*} 
This mismatch measure captures the discrepancy between the model output and observational data that is common across all input parameter settings and translate it into the logit scale. The simulation study in \cite{chang2015binary} shows that this discrepancy vector gives a parsimonious and reasonably accurate representation of discrepancy when the design points $\btheta_1,\dots,\btheta_n$ are representative sample of possible values of $\btheta^*$.

\subsection{Bayesian Inference} \label{sec:BayesianInference}

Given the above formulation we conduct Bayesian inference on $\btheta^*$ and other parameters in the model. While using non-Bayesian inference might be possible as well, we choose to use a Bayesian method as it provides a quite straightforward way to quantify the uncertainty in $\btheta^*$ while account for other sources of uncertainties  despite the complexity of our model specification.

\paragraph*{Likelihood}

In a similar fashion to the specification in \eqref{equation:ZbasicMixture} the representations in \eqref{equation:BasisRepForZ} and \eqref{equation:BasisRepForLambda} lead to a density function based on a mixture model. The likelihood function for the mixture model conditional on the emulated process $\bxi$ and $\bpsi$  now becomes
\begin{align*}
  f\left( Z(\bs_j)|\bxi(\btheta^*),\mathbf{r},\sigma^2_\epsilon,\bpsi(\btheta^*),\bv \right)
    =& \left| \frac{\partial Z(\bs_j) }{\partial t(\bs_j)}\right|  f\left(Z(\bs_j)|\bxi(\btheta^*),\mathbf{r},\sigma^2_\epsilon\right)P\left(I_z(\bs_j)=1|\bpsi(\btheta^*),\bv\right) \\
    &+\mathcal{D}_0\left(Z(\bs_j)\right)P\left(I_z(\bs_j)=0|\bpsi(\btheta^*),\bv\right) 
\end{align*}  
for all locations $\bs_1,\dots,\bs_p$, where the density function  $f\left(\cdot|\bxi(\btheta^*),\mathbf{r},\sigma^2_\epsilon \right)$ is for the case with $I_z(\bs_j)=1$ in \eqref{equation:BasisRepForZ}. As a result we have the following likelihood function for $\bZ$:
\begin{align*}
\mathcal{L}( \bZ|\bxi(\btheta^*),\mathbf{r},\sigma^2_\epsilon,\bpsi(\btheta^*),\bv)
\propto & \prod_{j=1}^{m} f\left(Z(\bs_j)|\bxi(\btheta^*),\mathbf{r},\sigma^2_\epsilon\right)P(I_z(\bs_j)=1|\bpsi(\btheta^*),\bv)\\
& \times \prod_{j=m+1}^{p} P(I_z(\bs_j)=0|\bpsi(\btheta^*),\bv).\\
= & \mathcal{L}_1\left(\bZ^{+}|\bxi(\btheta^*),\mathbf{r},\sigma^2_\epsilon\right) \mathcal{L}_2(I_z(\bs_1),\dots, I_z(\bs_p)|\bpsi(\btheta^*),\bv),
\end{align*}  
where 
\begin{align*}
\mathcal{L}_1\left(\bZ^{+}|\bxi(\btheta^*),\mathbf{r},\sigma^2_\epsilon\right)=&\prod_{j=1}^{m} f\left(Z(\bs_j)|\bxi(\btheta^*),\mathbf{r},\sigma^2_\epsilon\right),\\
\mathcal{L}_2(I_z(\bs_1),\dots, I_z(\bs_p)|\bpsi(\btheta^*),\bv)=&\prod_{j=1}^{m} P(I_z(\bs_j)=1|\bpsi(\btheta^*),\bv) \prod_{j=m+1}^{p} P(I_z(\bs_j)=0|\bpsi(\btheta^*),\bv).
\end{align*} 
We have a similar factorization as in \eqref{equation:FactoredLikelihoodForZ} with one factor for the positive observations $\bZ^{+}$ and the other for the binary variables at all locations $I_z(\bs_1),\dots, I_z(\bs_p)$.

\paragraph*{Prior}
 
To complete the Bayesian model, we assign the following priors for the model parameters ($\btheta^*,\bv,\sigma_{r}^{2},\sigma_{\epsilon}^{2},\sigma_{v}^{2},\boldsymbol{\kappa}_{u}$, and $\bR$) in our calibration step:
\begin{align*}
v_{j}|\sigma_{v}^{2} & \sim N(0,\sigma_{v}^{2})\;j=1,...,J_{v}\:; & \sigma_{v}^{2} & \sim IG(2,1)\:;\\
\sigma_{r}^{2} & \sim IG(2,3)\:; & \sigma_{\epsilon}^{2} & \sim IG(10,11000)\\
\kappa_{u,j} & \sim IG(5,6\hat{\kappa}_{u,j})\;j=1,...,J_{u}\:; & \mathbf{R} & \sim f(\mathbf{R})\\
\btheta^* & \sim d\pi(\btheta^*)
\end{align*}
 where $f(\mathbf{R})$ is a uniform distribution within the
range that $\mathbf{I}_{J_{r}}-\mathbf{R}\mathbf{R}^{T}$ is positive
definite, i.e., $f(\mathbf{R})\propto I(\mathbf{I}_{J_{r}}-\mathbf{R}\mathbf{R}^{T}\mbox{is positive definite})$, and $\pi(\btheta^*)$ is the uniform distribution defined over the physically possible range for the parameters $\btheta^*$. Notice that we have specified weakly informative priors on $\sigma_{v}^{2}$
and $\sigma_{r}^{2}$ for computational stability reasons however
we have noticed in our pilot simulations that the posterior analysis
is insensitive to the choice of the prior hyper-parameters. For $\sigma_{\epsilon}^{2}$
, we assigned moderately informative prior with purpose to encourage
$\sigma_{\epsilon}^{2}$ to take a value of around 1000. For the re-estimated
partial sill parameters $\kappa_{u,1},\dots,\kappa_{u,J_{u}}$, we
assigned a slightly informative prior to encourage them to have values
around their MLEs estimated in the emulation stage. To account uncertainty
on the input parameters $\btheta^*$, we assign independent uniform
priors within $[0,1]$ range on the input parameters $\btheta^*$ because
we have already re-scaled the parameter values in the range $[0,1]$,
whose limits in the original scale correspond to the physically possible
ranges of the input parameters.

\paragraph*{Posterior}

The above specification of likelihood and prior lead to a posterior whose density can be factorized as follows: 
\begin{align} 
\pi(\btheta^*,\bv,\sigma_{r}^{2},\sigma_{\epsilon}^{2},\sigma_{v}^{2},\boldsymbol{\kappa}_{u},\bR|\mathbf{Z})\propto & \pi_{1}(\btheta^*,\sigma_{r}^{2},\sigma_{\epsilon}^{2},\boldsymbol{\kappa}_{u},\bR|\bv,\bZ^{+})\label{eq:jointposterior}\\
 & \times\pi_{2}(\btheta^*,\bv,\sigma_{v}^{2}|I_{z}(\bs_{1}),\dots,I_{z}(\bs_{p})).\nonumber 
\end{align}
The first part on the right-hand side is based on the likelihood for
$\bZ^{+}$ ($\mathcal{L}_{1}$) and the relevant priors and obtained by
\begin{align*}
\pi_{1}(\btheta^*,\sigma_{r}^{2},\sigma_{\epsilon}^{2},\boldsymbol{\kappa}_{u},\bR|\bv,\bZ^{+})\propto & \int\int\mathcal{L}_{1}\left(\bZ^{+}|\bxi(\btheta^*),\mathbf{r},\sigma_{\epsilon}^{2}\right)f(\bxi(\btheta^*)|\btheta^*,\boldsymbol{\kappa}_{u})f(\br|\sigma_{r}^{2},\bv)d\mathbf{r}d\bxi\\
 & \times f(\btheta^*)f(\sigma_{r}^{2})f(\boldsymbol{\kappa}_{u})f(\sigma_{\epsilon}^{2})f(\bR)\\
= & \mathcal{L}_{1}^{*}(\bZ^{+}|\btheta^*,\sigma_{r}^{2},\sigma_{\epsilon}^{2},\boldsymbol{\kappa}_{u},\bv,\bR)f(\sigma_{r}^{2})f(\btheta^*)f(\boldsymbol{\kappa}_{u})f(\sigma_{\epsilon}^{2})f(\bR),
\end{align*}where $f(\sigma_{r}^{2})$, $f(\btheta^*)$, $f(\boldsymbol{\kappa}_{u})$,
$f(\sigma_{\epsilon}^{2})$ and $f(\bR)$ are the prior densities
(defined below) and the marginal likelihood $\mathcal{L}_{1}^{*}$
can be written as 
\begin{align*}
\mathcal{L}_{1}^{*}(\bZ^{+}|\btheta^*,\sigma_{r}^{2},\sigma_{\epsilon}^{2},\boldsymbol{\kappa}_{u},\bv,\bR)\propto\left|\Sigma_{+}\right|^{-1/2}\exp\left[-\frac{1}{2}(q^{-1}(\bZ^{+})-\bmu_{+})^{T}\Sigma_{+}^{-1}(\btheta^*)(q^{-1}(\bZ^{+})-\bmu_{+})\right],
\end{align*}
with $q^{-1}(\bZ^{+})=[q^{-1}(Z(\bs_{1})),\dots,q^{-1}(Z(\bs_{m}))]^{T}$.
The mean and covariance of $q^{-1}(\bZ^{+})$ are given by 
\begin{equation} \label{equation:SigmaForPositive}
\begin{aligned}
\bmu_{+} & =\bK_{+,u}\bmu_{\xi}(\btheta^*)-\bK_r\bmu_{\br|\bv}\\
\Sigma_{+} & =[\mathbf{K}_{+,u}~\mathbf{K}_r]\Sigma_{\xi,r}[\mathbf{K}_{+,u}~\mathbf{K}_r]^{T}+\sigma_{\epsilon}^{2}\mathbf{I}_m.
\end{aligned}
\end{equation}
Here $\bmu_{\xi}(\btheta^*)$ is the mean of the emulated process $\bxi(\btheta^*)$
and $\mathbf{K}_{+,u}$ is an $m\times J_{y}$ matrix created by collecting
the first $m$ rows of $\mathbf{K}_{u}$; $\Sigma_{\xi,r}$ is a block diagonal matrix
defined as 
\[
\Sigma_{\xi,r}=\left(\begin{array}{cc}
\Sigma_{\xi} & \mathbf{0}\\
\mathbf{0} & \Sigma_{\br|\bv}
\end{array}\right),
\]
where $\Sigma_{\xi}$ is a $J_{u}\times J_{u}$ diagonal matrix whose diagonal
elements are the conditional variances of $\xi_{1}(\btheta^*),\dots,\xi_{J_{u}}(\btheta^*)$
from the GP emulators defining $\bxi(\btheta^*)$; $\mu_{\br|\bv}$
and $\Sigma_{\br|\bv}$ are the conditional mean and variance of $\br$
given $\bv$ defined as 
\begin{align*}
\bmu_{\br|\bv} & =\frac{\sigma_{r}}{\sigma_{v}}\mathbf{R}\mathbf{v},\\
\Sigma_{\br|\bv} & =\sigma_{r}^{2}\left(\mathbf{I}_{J_{r}}-\mathbf{R}\mathbf{R}^{T}\right).
\end{align*}
The computational cost for finding the inverse and the determinant
of this covariance matrix can be significantly reduced using the Sherman-Woodbury-Morrison
formula \citep{woodbury1949stability} and the determinant formula \citep{harville2008matrix}. See Appendix
A for further details.

\noindent 

The second part of the posterior density is given as 
\[
\pi_{2}(\btheta^*,\bv,\sigma_{v}^{2}|I_{z}(\bs_{1}),\dots,I_{z}(\bs_{p}))\propto\mathcal{L}_{2}(I_{z}(\bs_{1}),\dots,I_{z}(\bs_{p})|\bpsi(\btheta^*),\bv)f(\bpsi(\btheta^*)|\btheta^*)f(\bv|\sigma_{v}^{2})f(\sigma_{v}^{2}),
\]
with the prior densities $f(\bpsi(\btheta^*)|\btheta^*)$, $f(\bv|\sigma_{v}^{2})$,
and $f(\sigma_{v}^{2})$. The formulation for the logits in \eqref{equation:BasisRepForLambda}
leads to the following likelihood function for $I_{z}(\bs_{1}),\dots,I_{z}(\bs_{p})$:
\[
\mathcal{L}_{2}(I_{z}(\bs_{1}),\dots,I_{z}(\bs_{p})|\bpsi(\btheta^*),\bv)\propto\prod_{j=1}^{p}\left[\left(\frac{\exp(\lambda(\bs_{j}))}{1+\exp(\lambda(\bs_{j}))}\right)^{I_{z}(\bs_{j})}\left(\frac{1}{1+\exp(\lambda(\bs_{j}))}\right)^{1-I_{z}(\bs_{j})}\right],
\]
where the logits $\lambda(\bs_{1}),\dots,\lambda(\bs_{p})$ are determined
by $\bpsi(\btheta^*)$ and $\bv$ through the basis representation in
\eqref{equation:BasisRepForLambda}. The prior for $\bpsi(\btheta^*)$
is given by the Gaussian process emulator with the mean and covariance
respectively given in \eqref{equation:covariance_for_u} and \eqref{equation:EmulMean}
and hence has the following multivaraite normal density: 
\begin{align*}
f(\bpsi(\btheta^*)|\btheta^*)\propto\left|\Sigma_{\psi}(\btheta^*)\right|^{-\frac{1}{2}}\exp\left(-\frac{1}{2}[\bpsi(\btheta^*)-\bmu_{\psi}(\btheta^*)]^{T}\Sigma_{\psi}^{-1}(\btheta^*)[\bpsi(\btheta^*)-\bmu_{\psi}(\btheta^*)]\right),
\end{align*}
where $\bmu_{\psi}(\btheta^*)$ is a vector of conditional means given
by the Gaussian process emulators for $\mathbf{u}(\btheta^*)$; $\Sigma_{\psi}(\btheta^*)$
is a diagonal matrix whose diagonal elements are given by the conditional
variance from the same Gaussian process emulators.

The target input parameters and the other parameters can be inferred based on the posterior density in \eqref{eq:jointposterior}. To facilitate the Bayesian inference we can resort to MCMC methods, which, in our case, require sampling from the posterior distribution by using Metropolis within Gibbs sampling \citep{gilks1995markov,gelfand1990sampling}.

\subsection{Procedure Summary}

We conclude this section by summarizing the overall steps of our approach. Given the $n\times p$ matrix for model output $\bY$ and $p$-dimensional vector for observational data $\bZ$,

\begin{enumerate}
    \item Create a $n\times p$ matrix for ice-no ice binary patters, $\left\{ I_y(\btheta_i,\bs_j) \right\} (i=1\dots,n,j=1,\dots,p)$, by dychotomizing the elements in the model output matrix $\bY$ into 0s and 1s. Apply LPCA to the dychotomized output matrix to find the $n\times J_w$ matrix for LPC scores $\bW$. 
    \item Apply likelihood-based PCA only to the non-zero values in $\bY$, to find   the $n\times J_u$ matrix for PC scores $\bU$. 
    \item For each column in $\bW$ and $\bU$, separately construct a 1-dimensional GP emulator by finding MLEs for the emulator parameters. Let $\bpsi(\btheta)$ and $\bxi(\btheta)$ respectively denote $J_w$- and $J_u$-dimensional emulated processes for the unobserved values of $\bw(\btheta)$ and $\bu(\btheta)$, which are collections of independently constructed 1-dimensional GP emulators.
    \item Infer the best input parameter setting $\btheta^*$ along with other parameters based on the posterior density given the observational data $\bZ$ (see Equation \eqref{eq:jointposterior} for its definition). The Bayesian inference can be facilitated through Metropolis-within-Gibbs sampling.
\end{enumerate}

\section{Application}
\label{sec:Results}

We now discuss the results of applying our method to the problem of PSU-3D ice model calibration based on Bedmap2 data described in Section \ref{sec:ModelRunsAndObs}. As the first step, we have built a dimension-reduced emulator described in Section \ref{sec:DimReduction}, which takes about 5 hours on a single high-performance core if implemented in an R code. While further speed-up is possible by switching to a faster programming language or utilizing parallel computing we have decided not to pursue such an effort as the current implementation is fast enough for our purpose. We use 10 LPCs (i.e. $J_w=10$) and 20 PCs (i.e. $J_u=20$) as using more PCs does not yield meaningful improvement in emulation performance. 

\begin{figure}
\centering
\includegraphics[scale=0.50]{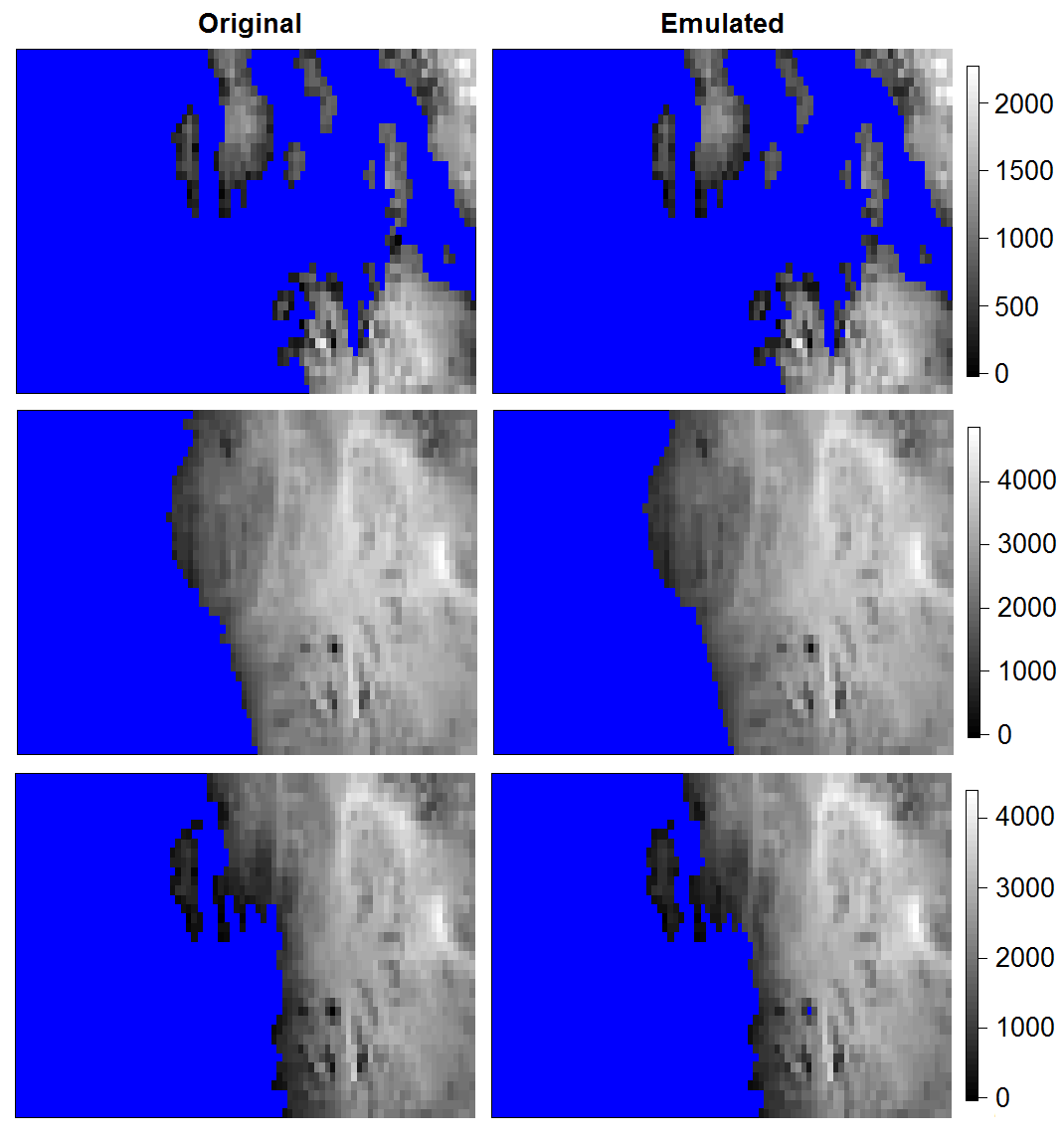}
\caption{Examples of leave-10\%-out validation results, showing selected original spatial patterns from PSU-3D ice model (left column) and the corresponding emulated patterns (right column). The comaprison shows that our emulator can predict the original model output with high accuracy.}
\label{fig:leave10out}
\end{figure}

To verify the performance of our emulator we first conduct leave-10\%-our cross-validation for the emulator: i.e., we have randomly left out 10\% of the model runs and tried to predict them using the constructed emulator. Some example cases are shown in Figure \ref{fig:leave10out}. The cross-validation results show that our emulator can predict the left-out model outputs with a high accuracy, both in terms of the ice-no ice binary patterns and the thickness patterns. The overall mean absolute error (MAE) for ice thickness prediction at the locations with positive thickness is about 96m (while the overall mean ice thickness at those locations is 2117m). The sensitivity (the percentage of left out runs where ice presence was correctly predicted) is 98.5\% and the specificity (the percentage of left out runs where ice absence was correctly predicted) is 96.1\%.

Using the constructed emulators and  the observational data we infer the best input parameter setting $\btheta^*$. We first verify our method using a synthetic data example in Section \ref{sec:ResultsSynthetic} and proceed to calibration using the real observations from Bedmap2 data in Section \ref{sec:ResultsRealObs}. In both cases we compare our current method (full approach henceforth) to the calibration results obtained using only the ice/no ice binary patterns \citep[binary-only approach henceforth, originally presented in][]{chang2015binary} to show the added value by fully utilizing the ice thickness patterns in calibration.

\subsection{Choice of Transformation Function and Emulation Performance}

The success of this latent variable-based approach partially depends on the choice of the transformation function $q$ to guarantee non-negativity without introducing a serious artifact due to transformation. While in the literature an exponential transformation is commonly used to enforce non-negativity, we found that the use of an exponential transformation imposes too much distortion in distribution and results in a poor emulation performance in our problem (MAE of about 401m, four times higher than that of our result). Therefore in this study we use the following link function that can ensure non-negativity with only a minimal distortion of data distribution:
\begin{equation*}
q(x)=
\begin{cases}
    x &,~ \mbox{if}~ x>1, \\
    \exp (x-1) &, ~\mbox{if}~ x\le 1.
      \end{cases}
\end{equation*}
This function preserves the original pattern of ice thickness by setting $h(\btheta_i,\bs_j)=Y(\btheta_i,\bs_j)$ for $Y(\btheta_i,\bs_j)>1$ m, while allowing the transformed variable can have negative values by setting $h(\btheta_i,\bs_j)=\log(Y(\btheta_i,\bs_j))+1$ for $0 < Y(\btheta_i,\bs_j) \le 1$ m. This function also ensures a smooth transition at $x=1$ because $\frac{ \partial q(x)}{\partial x} $ exists and has a value of one when $x=1$.

One drawback of the above transformation is that the calibration of the ice thickness $q(\eta(\theta^{*},\bs)+\delta(\bs)+\epsilon)$ is different for ice thickness smaller than one meter and for ice thickness greater than one meter. More precisely, the calibration formulation  is multiplicative  for ice thickness of magnitude less than one meter and  additive for ice thickness of greater or equal to one meter.  However, interesting observation regarding the WAIS application is that the percentage of ice thickness lower than one matter is practically zero. In our application we found that ice thickness of less than one meter amounts for $0.01\%$ of the total ice thickness  for both simulated and observed  data sets. This implies that our calibration process is in practice an additive calibration model.

\begin{figure}
\centering
\includegraphics[scale=0.7]{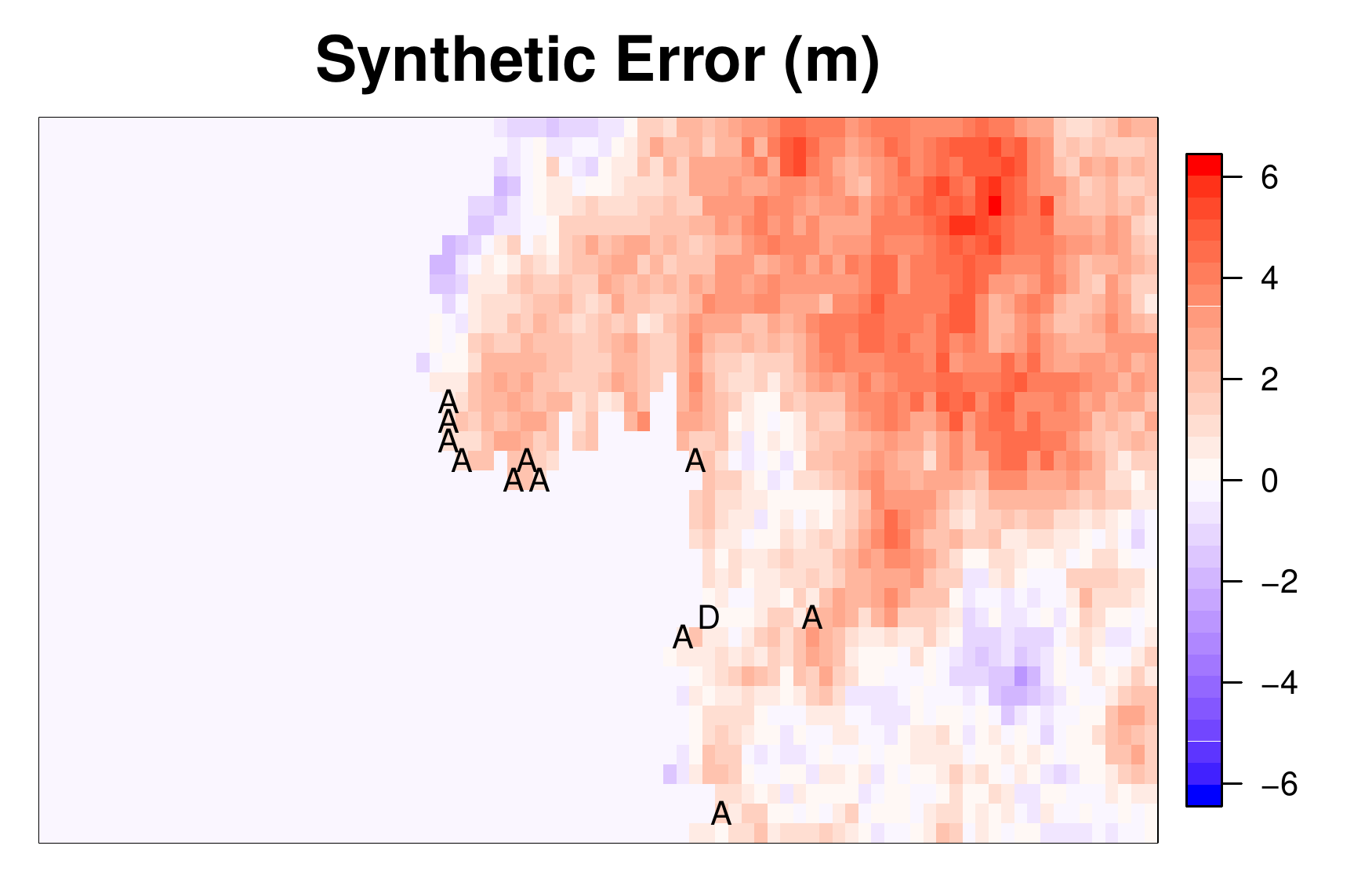}
\caption{Synthetic error generated as described in Section \ref{sec:ResultsSynthetic}. 'A' represents the location where ice is added (i.e., $I_z(\bs)$ is changed from 0 to 1) and 'D' represents the locations where existing ice is removed (i.e. $I_z(\bs)$ is changed from 1 to 0).}
\label{fig:SimulatedError}
\end{figure}

\subsection{Calibration Using Synthetic Data} \label{sec:ResultsSynthetic}
We now verify the performance of our calibration method using a synthetic data example. To generate a synthetic data set we choose the true input parameter setting and its corresponding output for ice thickness pattern as the assumed truth. We then superimpose generated errors to represent a possible data-model discrepancy in reality. We chose a model output whose input parameter values are not at the center of the cloud of design points to make the test more challenging. To create a synthetic ice/no ice binary pattern we have first chosen the top 30\%  model runs that are closest to the synthetic truth based on the mean squared error in thickness, and then calculated the average difference in thickness for each location and subtracted it from the synthetic true thickness. By dichotomizing the resulting thickness pattern we have obtained a `contaminated' binary pattern as shown in Figure \ref{fig:SimulatedError}. The same approach to obtaining a contaminated binary pattern was also used in \cite{chang2015binary}. For the discrepancy in ice thickness we have generated a pattern from a spatial Gaussian process model with an isotropic exponential covariance with the partial sill of 4m$^2$, the range of 400km, and the nugget of 0.01m$^2$. This represents a situation where the model is highly accurate in representing the modern ice thickness (and hence the data-model discrepancy is small) and the pattern of discrepancy has a long-range dependence. Figure \ref{fig:SimulatedError} shows the resulting error pattern for ice thickness. We have avoided a simpler approach of adding a random noise to the thickness pattern and dichotomizing the resulting pattern, because such approach tend to add extra locations with positive ice thickness too easily (because any `no ice' location would be turned into an `ice covered location' whenever a positive error value is added).

For both the full and the binary-only approaches, we respectively have obtained an MCMC chain with a length of 150,000 iterations and verified that it has reached equilibrium by comparing the first half and the whole MCMC chain (results not shown). The overall computing time has taken about $96$ hours on a high-performance single core with an R code implementation. Switching to a faster program language and applying parallelization will certainly make the computation much faster, but we did not seek such speed up here because the application problem at hand does not require a faster solution. To verify the performance of our method in terms recovering the assumed true input parameter setting we compare the estimated posterior densities with the assumed true input parameter settings. The results in Figure \ref{fig:PosteriorDensitySynthetic} show that the full approach can recover the assumed truth with a reasonable accuracy and yield sharper posterior densities compared to those on the binary-only approach. To confirm that the sharper posterior densities for input parameters by the full approach also result in better future projections we transform the MCMC sample for input parameters into a sample for the projected ice volume changes in 500 years, using another Gaussian process emulator constructed using the existing model runs described in Section \ref{sec:ModelRunsAndObs}. The resulting predictive distribution for the future projections in Figure \ref{fig:PredictiveSynthetic} show similar results: the method that fully utilizes the thickness patterns leads to sharper WAIS volume change predictions compared to the method only based on the binary patters.

\subsection{Calibration Using Real Observational Data } \label{sec:ResultsRealObs}
We now apply our calibration approach to the Bedmap2 dataset introduced in \ref{sec:ModelRunsAndObs}. The resulting estimated posterior density for the input parameters is illustrated in Figure \ref{fig:PosteriorDensityObs}. As in Section \ref{sec:ResultsSynthetic} we compare the results based only on the binary patterns and those based on the full thickness patterns. Similarly to the synthetic data example utilizing the information from ice thickness makes the posterior density for the input parameters sharper, by ruling out parameter settings that create a similar binary pattern to the observational data but has a very different ice thickness pattern. We also observe that the bivariate marginal densities exhibit some bimodality except for the joint density plot for OCFAC and CRH. This seems to be due to the fact that two different kinds of combinations, a middle range value of TAU and a higher value of CALV or a lower value of CALV and a higher value of TAU, lead to equally good simulation results for ice thickness. We note that understanding the interactions between the input parameters based only on these marginal density plots is not easy and further careful investigation is required to fully understand the possibly complicated identifiability issues for input parameters. The predictive density for ice volume change is also sharpened with a similar peak but slightly shifted overall distribution towards smaller values. This shift towards the left is mainly due to the fact that higher values for OCFAC is ruled out by the information from ice thickness.

\section{Summary and Future Directions}
\label{sec:Discussion}

In this paper we have formulated an efficient emulation and calibration method that can handle semi-continuous spatial model output and observational data, which often arise in scientific fields such as glaciology and meteorology. We use a mixture  model for the semi-continuous output which results in a multiplicative  representation of the likelihood between the binary and continuous part of the dataset. Using dimension reduction and basis representation techniques, our approach can overcome the inferential and computational challenges posed by high-dimensional and dependent semi-continuous data and  provide a statistically sound way to quantify input parameter uncertainties. In a simulation setting, we have shown that our approach can recover the true input parameter values and lead to smaller parametric and prediction uncertainties when compared to methods that aggregate or simplify the observations and model output, say by converting the semi-continuous data into binary data. Similar reduction in parametric and prediction uncertainties are also observed in the real data example with the Bedmap2 dataset. We have demonstrated the value of our approach in the context of a well known model for the Antarctic ice sheet. The methodology we have described here can also be applied to a wide range of calibration problems that involve semi-continuous spatial or image data. In the field of climate science and meteorology, for example, many important processes such as precipitation, pollution, and storm surge level are in the form of semi-continuous spatial data. 

Possible extensions of the proposed approach are as follows: First, our approach can be easily modified and applied to an application problem that involves model output and observational data in the form of zero-inflated count spatial data. Such data often arise in ecology applications, where the subjects of study such as animal or plant species show zero prevalence in a large portion of the study area. Second, our approach models the binary patterns indirectly through the logit. This forces to define a specific type of `nugget' effect defined by the marginal Bernoulli distribution at each location. Relaxing this assumption will lead to a more flexible model specification. Finally, one can modify our approach for using spatio-temporal model output and observational data. Such extension may require accounting for some complicated temporal dependence or even spatio-temporal interactions, as well as more serious data size issues. 

\section*{Acknowledgement}
This material was based upon work partially supported by the National Science Foundation under Grant DMS-1638521 to the Statistical and Applied Mathematical Sciences Institute. Any opinions, findings, and conclusions or recommendations expressed in this material are those of the authors and do not necessarily reflect the views of the National Science Foundation.

\appendix                                     
\section*{Appendix A: Matrix Computation in Section 4.3}
Let $\mathbf{K}_{+}=[\mathbf{K}_{+,u}~\mathbf{K}_{r}]$, then the covariance matrix in \eqref{equation:SigmaForPositive} can be rewritten as
\begin{equation*}
\begin{aligned}
\Sigma_{+} & =[\mathbf{K}_{+,u}~\mathbf{K}_{r}]\Sigma_{\xi,r}[\mathbf{K}_{+,u}~\mathbf{K}_{r}]^{T}+\sigma_{\epsilon}^{2}\mathbf{I}_{m}.\\
&= \mathbf{K}_{+} \Sigma_{\xi,r} \mathbf{K}_{+}^T + \sigma_{\epsilon}^{2}\mathbf{I}_{m}.
\end{aligned}
\end{equation*}
By applying the Sherman-Morrison-Woodbury formula \citep{woodbury1949stability}, the inverse of this matrix can be expressed as
\begin{equation*}
(\mathbf{K}_{+} \Sigma_{\xi,r} \mathbf{K}_{+}^T + \sigma_{\epsilon}^{2}\mathbf{I}_{m})^{-1}=\sigma_{\epsilon}^{-2}\mathbf{I}_{m}-\sigma_{\epsilon}^{-2}\bK_+\left(\Sigma_{\xi,r}^{-1} + \sigma_{\epsilon}^{-2} \bK_+^T  \bK_+ \right)^{-1} \bK_+^T \sigma_{\epsilon}^{-2}.
\end{equation*}
This reduces the order of the computational cost of matrix inversion from $\mathcal{O}(n^3)$ to $\mathcal{O}(n^2)$. In a similar fashion, by applying the determinant formula \citep{harville2008matrix} the determinant of the matrix can be rewritten as
\begin{equation*}
\left| \mathbf{K}_{+} \Sigma_{\xi,r} \mathbf{K}_{+}^T + \sigma_{\epsilon}^{2}\mathbf{I}_{m} \right|= \sigma_{\epsilon}^{2m} \left|  \Sigma_{\xi,r}^{-1}+\sigma_{\epsilon}^{-2} \bK_+^T  \bK_+ \right| \left| \Sigma_{\xi,r} \right| 
\end{equation*}
This gives a similar computational gain as the Sherman-Morrison-Woodbury formula.


\begin{figure}[h]
\centering
\includegraphics[scale=0.43]{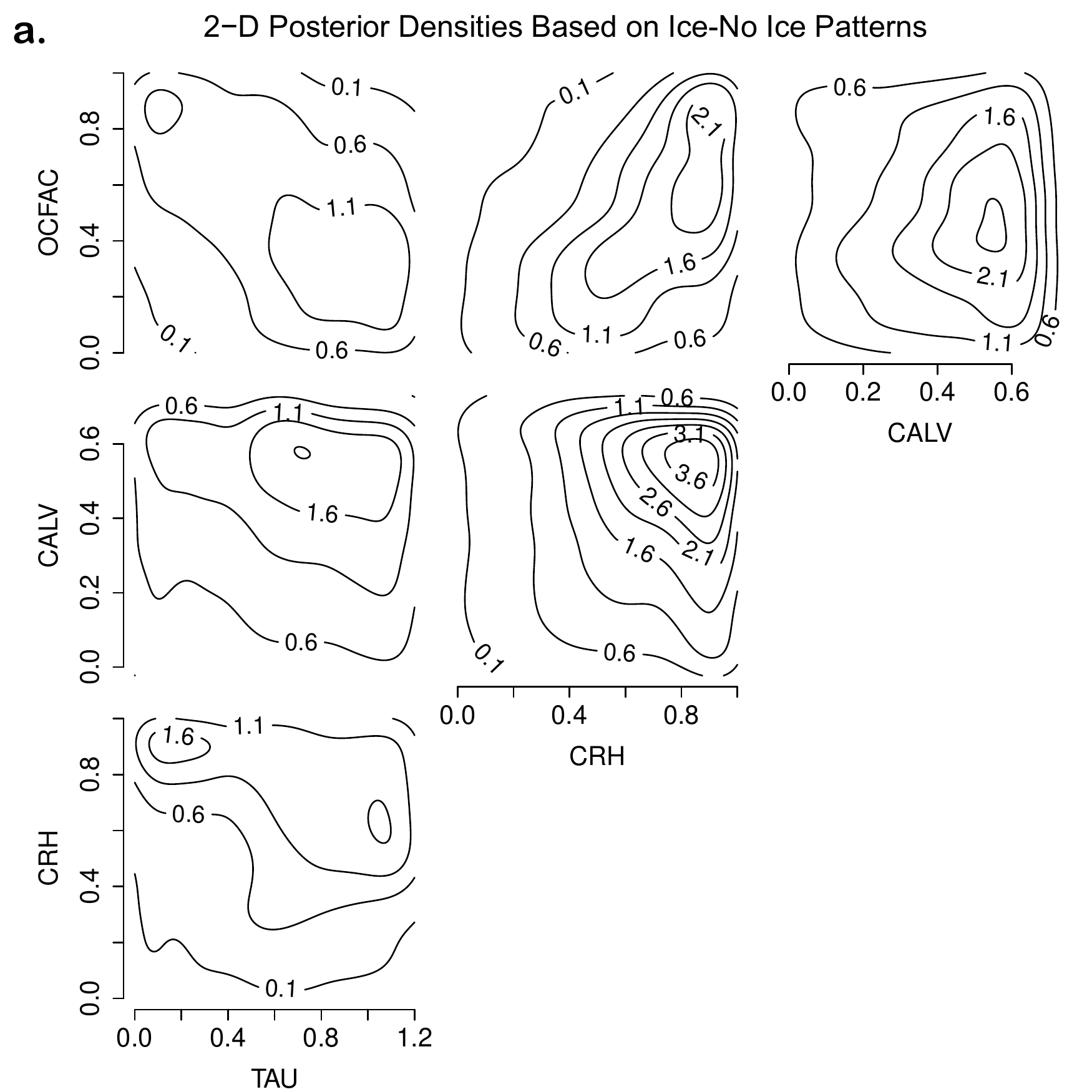}
\includegraphics[scale=0.43]{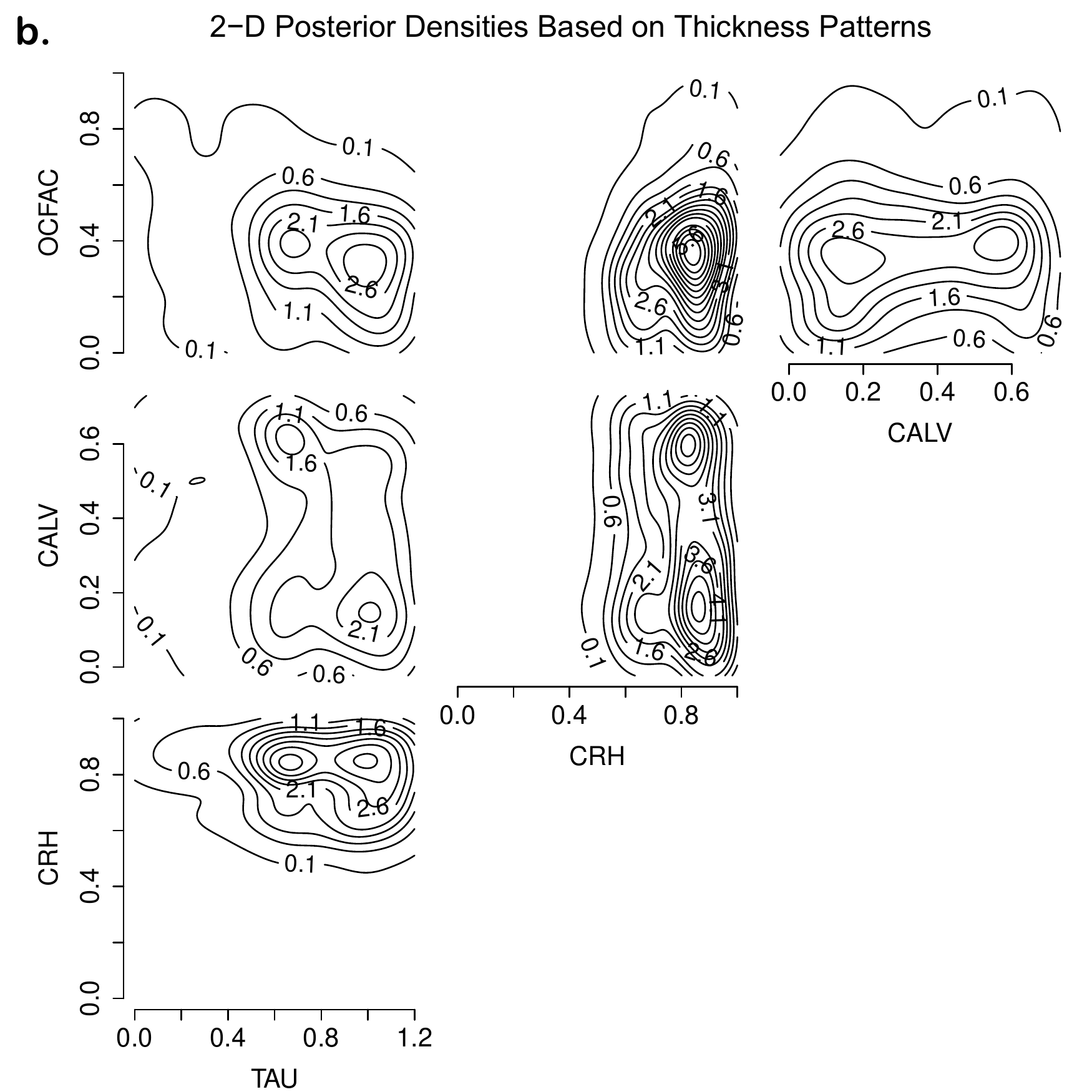}
\caption{Two-dimensional marginal densities of the input parameters for the real data example described in Section \ref{sec:ResultsRealObs}, estimated based only on the binary patterns (a) and the full ice thickness patterns (b). Again the values of input parameters are re-scaled as displayed in the x- and y- axes for easier presentation. Similarly to the results in Figure \ref{fig:PosteriorDensitySynthetic} the results based on the full ice thickness patterns leads to sharper densities.}
\label{fig:PosteriorDensityObs}
\end{figure}
\clearpage

\begin{figure}[h]
\centering
\includegraphics[scale=0.50]{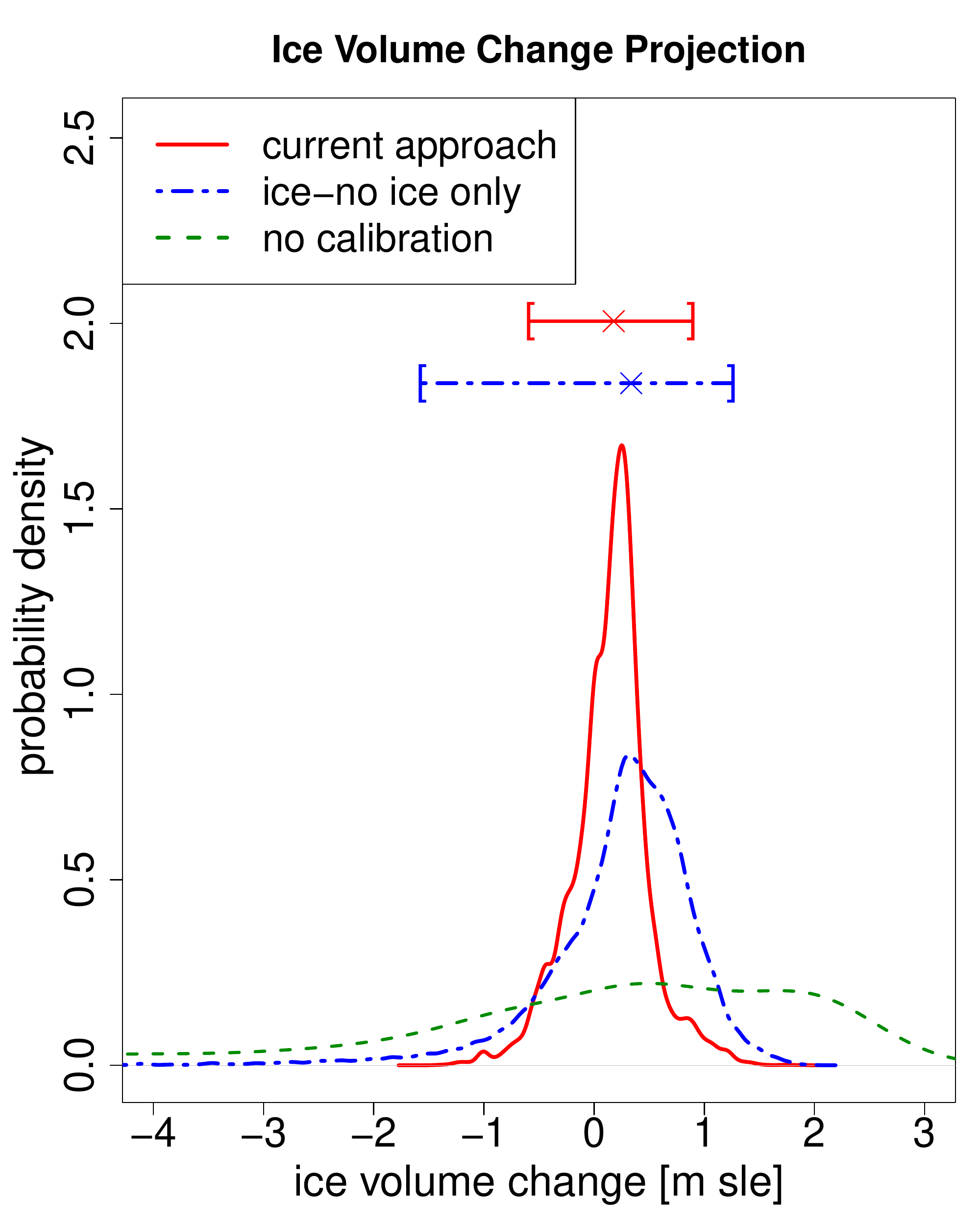}
\caption{The same as Figure \ref{fig:PredictiveSynthetic} except that the results are based on the densities in Figure \ref{fig:PosteriorDensityObs}, the posterior densities for observational data. Again, the projection based on the full ice thickness has the sharpest density.}
\label{fig:PredictiveObs}
\end{figure}
\clearpage

\begin{figure}[h]
\centering
\includegraphics[scale=0.43]{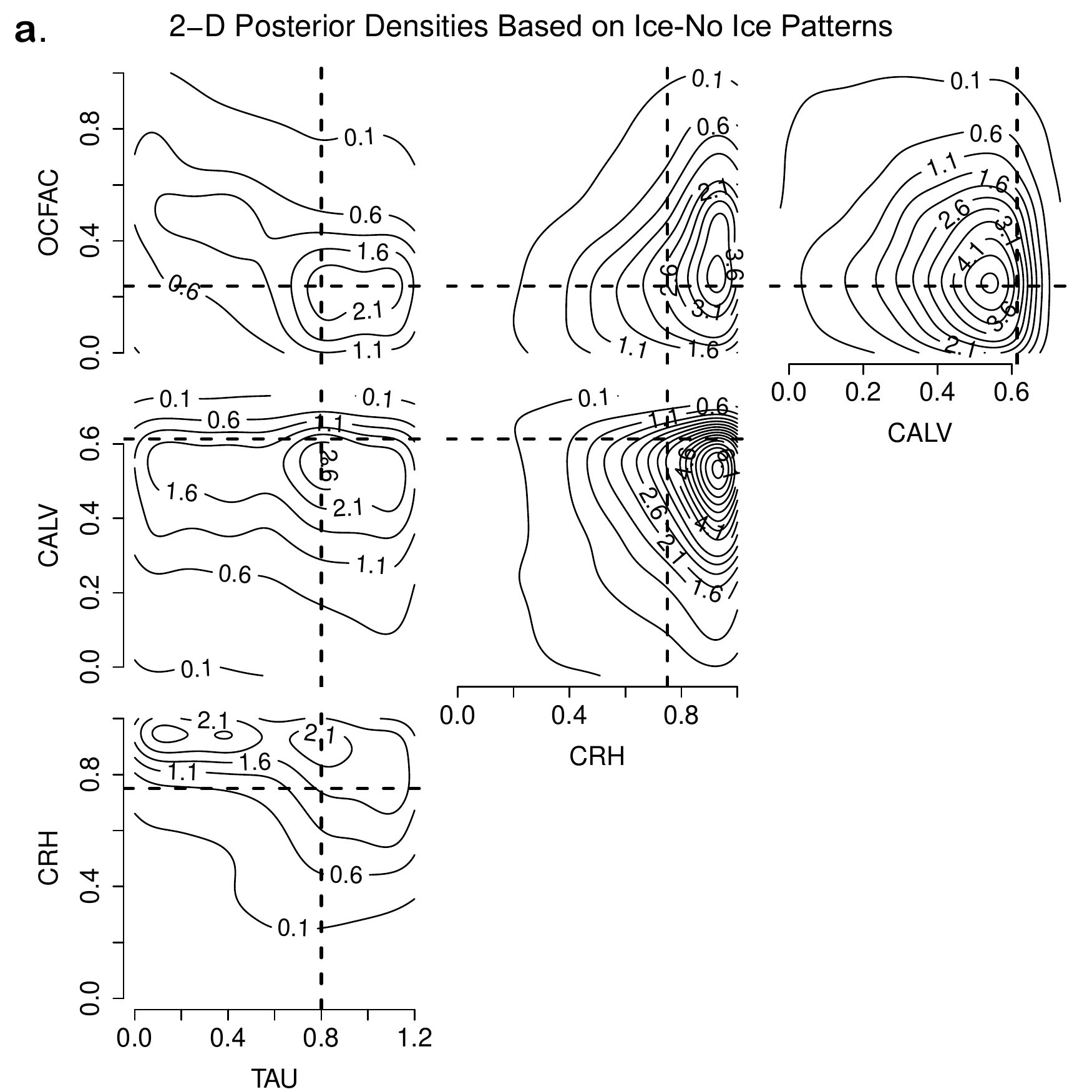}
\includegraphics[scale=0.43]{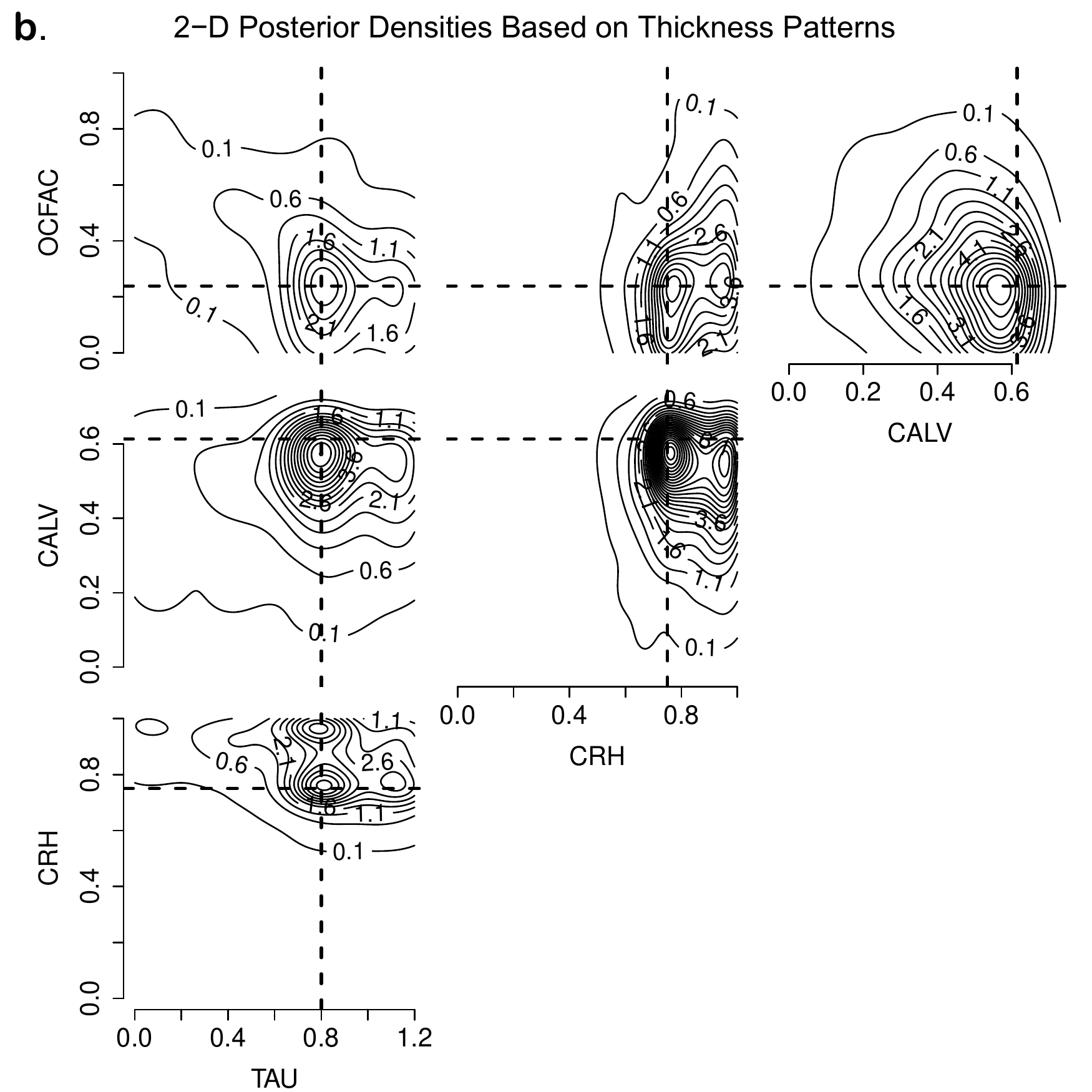}
\caption{ Two-dimensional marginal densities of the input parameters for the synthetic data example described in Section \ref{sec:ResultsSynthetic}, estimated based only on the binary patterns (a) and the full ice thickness patterns (b). The values of input parameters are re-scaled as shown in the axes for ease of presentation. While both densities are informative about the assumed true input parameter setting (shown as dashed lines), calibration based on the full ice thickness patterns yields  sharper densities.}
\label{fig:PosteriorDensitySynthetic}
\end{figure}
\clearpage

\begin{figure}[h]
\centering
\includegraphics[scale=0.50]{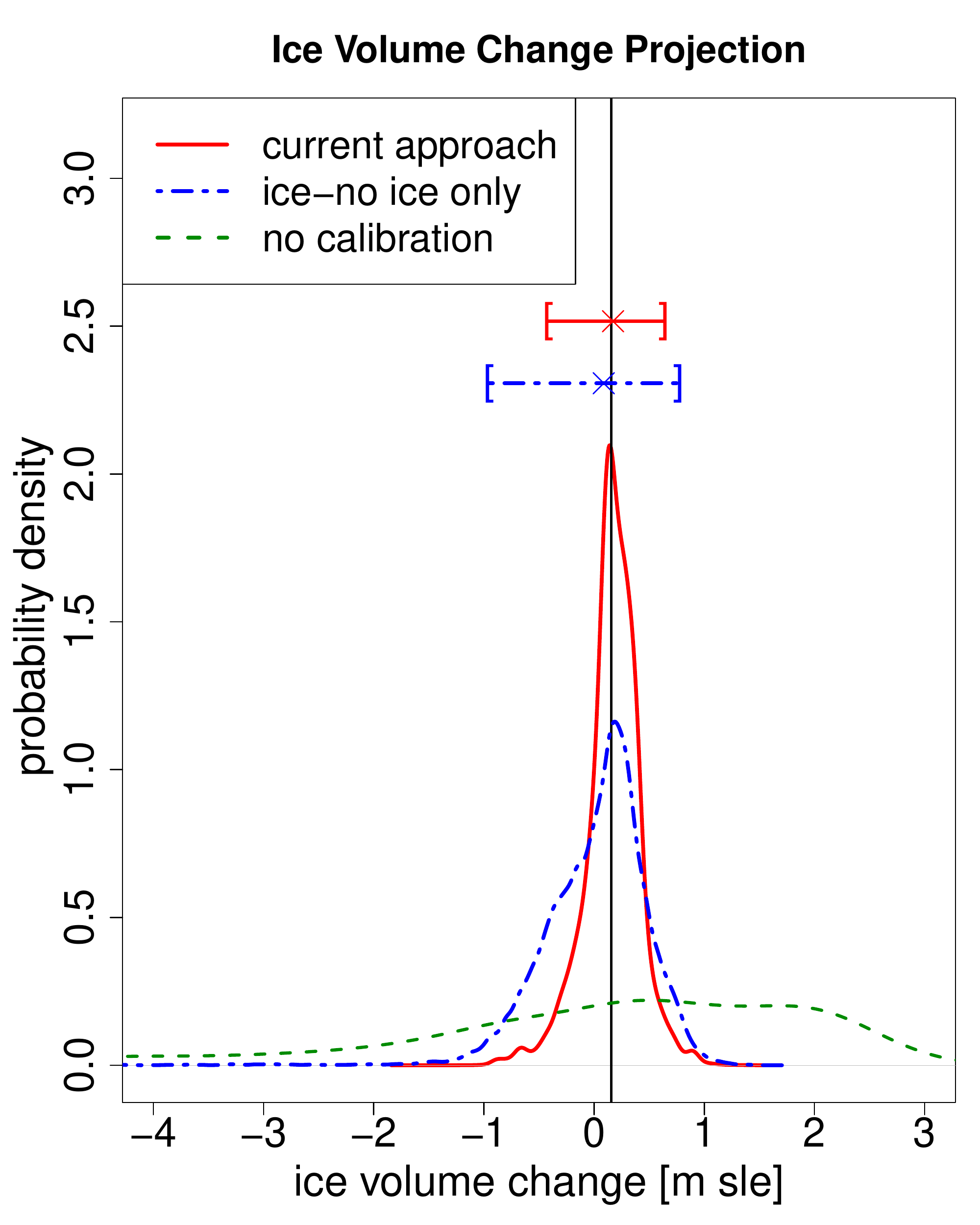}
\caption{Ice volume change projects based on the estimated posterior densities show in Figure \ref{fig:PosteriorDensitySynthetic}. The projection based on the full thickness patterns (solid line) has a  sharper density than that based on the binary patterns only (dashed and dotted line). The dashed line shows the projection density created by assigning a uniform density over the entire input parameter ranges. The modes of the densities from both results are close to the assumed true ice volume change projection (vertical solid line), but the projection density based on the full ice thickness patterns is a sharper than that based on the binary patterns only.}
\label{fig:PredictiveSynthetic}
\end{figure}

\end{document}